\documentclass[aps,prl,twocolumn,groupedaddress]{revtex4}

\usepackage[dvips]{graphicx,color} % Graphics and color
\usepackage{array,hhline,dcolumn} % Better table handling
\usepackage{rotating} % Rotate and sideways environments

\newcommand{\gtwid}{\mathrel{\raise.3ex\hbox{$>$\kern-.75em\lower1ex\hbox{$\sim$}}}}
\newcommand{\ltwid}{\mathrel{\raise.3ex\hbox{$<$\kern-.75em\lower1ex\hbox{$\sim$}}}}

\begin{document}
%\voffset=0.75in
%
%Below writes "DRAFT"
%\special{!userdict begin /bop-hook{gsave 200 30 translate
%65 rotate /Times-Roman findfont 216 scalefont setfont
%0 0 moveto 0.7 setgray (DRAFT) show grestore} def end}
% APS preprint designation
% \preprint{MiniBooNE-OSC}

%\title{Indication of Electron Antineutrino Appearance at the $\Delta m^2 \sim$ 1 $\mathrm{eV}^{2}$ Scale}
%\title{Indication of Electron Antineutrino Appearance Oscillations at the LSND Mass Scale}
%\title{Suggested $\bar \nu_\mu \rightarrow \bar \nu_e$ Oscillations at the 
%$\Delta m^2 \sim$ 1 $\mathrm{eV}^{2}$ Scale}
%\title{Further Evidence for $\bar \nu_\mu \rightarrow \bar \nu_e$ Oscillations at the 
%LSND Mass Scale}
%\title{Evidence for Electron Antineutrino Appearance Oscillations at the 
%LSND Mass Scale}
%\title{Excess Events in the Search for $\bar \nu_\mu \rightarrow \bar \nu_e$ Oscillations at the 
%$\Delta m^2 \sim$ 1 $\mathrm{eV}^{2}$ Scale}
\title{Event Excess in the MiniBooNE Search for 
$\bar \nu_\mu \rightarrow \bar \nu_e$ Oscillations}

\author{
        A.~A. Aguilar-Arevalo$^{12}$, C.~E.~Anderson$^{15}$, S.~J.~Brice$^{6}$,
        B.~C.~Brown$^{6}$, L.~Bugel$^{11}$, J.~M.~Conrad$^{11}$,
	R.~Dharmapalan$^{1}$, 
	Z.~Djurcic$^{2}$, B.~T.~Fleming$^{15}$, R.~Ford$^{6}$,
        F.~G.~Garcia$^{6}$, G.~T.~Garvey$^{9}$, J.~Mirabal$^{9}$,
        J.~Grange$^{7}$, J.~A.~Green$^{8,9}$,
        R.~Imlay$^{10}$,
        R.~A. ~Johnson$^{3}$, G.~Karagiorgi$^{11}$, T.~Katori$^{8,11}$,
        T.~Kobilarcik$^{6}$, S.~K.~Linden$^{15}$,
        W.~C.~Louis$^{9}$, K.~B.~M.~Mahn$^{5}$, W.~Marsh$^{6}$,
        C.~Mauger$^{9}$, 
        W.~Metcalf$^{10}$, G.~B.~Mills$^{9}$,
        C.~D.~Moore$^{6}$, J.~Mousseau$^{7}$, R.~H.~Nelson$^{4}$,
        V.~Nguyen$^{11}$, P.~Nienaber$^{14}$, J.~A.~Nowak$^{10}$,
        B.~Osmanov$^{7}$, Z.~Pavlovic$^{9}$, D.~Perevalov$^{1}$,
        C.~C.~Polly$^{6}$, H.~Ray$^{7}$, B.~P.~Roe$^{13}$,
        A.~D.~Russell$^{6}$, R.~Schirato$^{9}$,
	M.~H.~Shaevitz$^{5}$, M.~Sorel$^{5}$\footnote{Present address: IFIC, Universidad de Valencia and CSIC, Valencia 46071, Spain},
        J.~Spitz$^{15}$, I.~Stancu$^{1}$, R.~J.~Stefanski$^{6}$,
        R.~Tayloe$^{8}$, M.~Tzanov$^{4}$, R.~G.~Van~de~Water$^{9}$, 
        M.~O.~Wascko$^{10}$\footnote{Present address: Imperial College; London SW7 2AZ, United Kingdom},
        D.~H.~White$^{9}$, M.~J.~Wilking$^{4}$, G.~P.~Zeller$^{6}$,
        E.~D.~Zimmerman$^{4}$ \\
\smallskip
(The MiniBooNE Collaboration)
\smallskip
}
\smallskip
\smallskip
\affiliation{
$^1$University of Alabama; Tuscaloosa, AL 35487 \\
$^2$Argonne National Laboratory; Argonne, IL 60439 \\
$^3$University of Cincinnati; Cincinnati, OH 45221\\
$^4$University of Colorado; Boulder, CO 80309 \\
$^5$Columbia University; New York, NY 10027 \\
$^6$Fermi National Accelerator Laboratory; Batavia, IL 60510 \\
$^7$University of Florida; Gainesville, FL 32611 \\
$^8$Indiana University; Bloomington, IN 47405 \\
$^9$Los Alamos National Laboratory; Los Alamos, NM 87545 \\
$^{10}$Louisiana State University; Baton Rouge, LA 70803 \\
$^{11}$Massachusetts Institute of Technology; Cambridge, MA 02139 \\
$^{12}$Instituto de Ciencias Nucleares, Universidad Nacional Aut\'onoma de M\'exico, D.F. 04510, M\'exico \\
$^{13}$University of Michigan; Ann Arbor, MI 48109 \\
$^{14}$Saint Mary's University of Minnesota; Winona, MN 55987 \\
$^{15}$Yale University; New Haven, CT 06520\\
}

\date{\today}% It is always \today, today,
             %  but any date may be explicitly specified

\begin{abstract}
The MiniBooNE experiment at Fermilab reports results from a search for 
$\bar \nu_\mu \rightarrow \bar \nu_e$ oscillations,
using a data sample corresponding to $5.66 \times 10^{20}$ protons 
on target. An excess of $20.9 \pm 14.0$ events is observed in the
energy range $475<E_\nu^{QE}<1250$ MeV, which, when constrained
by the observed $\bar \nu_\mu$ events, 
has a probability for consistency
with the background-only hypothesis of 0.5\%. On the other hand,
fitting for $\bar{\nu}_{\mu}\rightarrow\bar{\nu}_e$ oscillations,
the best-fit point has a $\chi^2$-probability of 8.7\%.
%, while 
%the probability of the background-only fit relative to the best 
%oscillation fit is 0.6\%.
The data are consistent with $\bar \nu_\mu \rightarrow \bar \nu_e$ 
oscillations in the 0.1 to 1.0 eV$^2$ $\Delta m^2$ range and with the
evidence for antineutrino oscillations from the Liquid Scintillator 
Neutrino Detector at Los Alamos National Laboratory.
\end{abstract}

% PACS, the Physics and Astronomy
% Classification Scheme.

% Use showkeys class option if keyword
% display desired
\maketitle

% Include sections

The MiniBooNE experiment has published searches for 
$\nu_\mu \rightarrow \nu_e$ and $\bar \nu_\mu \rightarrow \bar \nu_e$
oscillations, motivated by the LSND $3.8 \sigma$ excess 
of $\bar \nu_e$ candidate 
events \cite{lsnd}. In the $\nu_\mu \rightarrow \nu_e$ 
study, MiniBooNE found
no evidence for an excess of $\nu_e$ candidate events above 475 MeV;
however, a $3.0 \sigma$ excess of electron-like events was 
observed below 475 MeV \cite{mb_osc,mb_lowe}.
The source of the excess remains unexplained \cite{mb_lowe}, although
several hypotheses have been put forward 
\cite{hhh,gninenko,karagiorgi,sterile,weiler,goldman,barger,akhmedov,nelson,kostelecky}, 
including, for example, anomaly-mediated neutrino-photon 
coupling, sterile neutrino decay, 
and sterile neutrino oscillations with CP or CPT violation. 
Initial results from the $\bar \nu_\mu \rightarrow \bar \nu_e$ study were 
reported in \cite{mb_osc_anti}. A search in antineutrino mode provides a more 
direct test of the LSND signal, which was observed with antineutrinos.
Due to limited statistics, the 
initial MiniBooNE $\bar \nu_\mu \rightarrow \bar \nu_e$ search was inconclusive with respect to 
two-neutrino oscillations at the LSND mass scale, although a joint 
analysis reported compatibility between the LSND, KARMEN \cite{karmen,compatibility}, and MiniBooNE antineutrino experiments \cite{karagiorgi}.
In this paper, we report an updated analysis of the 
$\bar \nu_\mu \rightarrow \bar \nu_e$ search with 1.7 times more 
protons on target (POT) than reported in \cite{mb_osc_anti}.

This analysis uses the same technique that was reported earlier \cite{mb_osc_anti} 
and assumes only $\bar \nu_\mu \rightarrow \bar \nu_e$ oscillations with no 
significant $\bar{\nu}_{\mu}$ disappearance and no $\nu_{\mu}$ oscillations. 
%Therefore, the oscillation search presented here is an explicit search for
%oscillations where CP or CPT is violated, or effectively violated.
In addition, no contribution from the observed neutrino mode low energy excess 
has been accounted for in the antineutrino prediction. These simplifications
may change the fitted $\bar \nu_\mu \rightarrow \bar \nu_e$ 
oscillation probability by a total of $\sim 10\%$.

The antineutrino flux is produced by 8 GeV protons from the Fermilab Booster
interacting on a beryllium target inside a magnetic focusing horn. 
Negatively charged mesons 
produced in p-Be interactions are focused in the forward direction 
and subsequently decay primarily into $\bar{\nu}_{\mu}$. The flux for 
neutrinos and antineutrinos of all flavors is calculated with a
simulation program using external measurements \cite{mb_flux}.
In antineutrino mode, the $\nu_\mu$, $\bar \nu_e$, and $\nu_e$ flux
contaminations at the detector are 15.7\%, 0.4\%, and 0.2\%,
respectively.
The $\bar{\nu}_{\mu}$ flux peaks at 400 MeV and has a mean energy of 600 MeV. 

The MiniBooNE detector has been described in detail elsewhere \cite{mb_detector}. The detector 
location was chosen to satisfy $L[\mathrm{m}]/E[\mathrm{MeV}]\sim1$, similar to that of LSND, which
maximizes the sensitivity to oscillations at $\Delta m^2\sim1$ eV$^{2}$. The detector consists of
a 40-foot diameter sphere filled with pure mineral oil ($\sim$CH$_{2}$). Neutrino interactions in the 
detector produce final-state electrons or muons, which produce scintillation and Cherenkov light 
detected by the 1520 8-inch photomultiplier tubes (PMTs) that line the interior of the detector. 

The signature of $\bar \nu_\mu \rightarrow \bar \nu_e$ oscillations 
is an excess of $\bar \nu_e$-induced charged-current quasi-elastic (CCQE) 
events. Reconstruction \cite{mb_recon} and selection requirements of these 
events are identical to those of the previous neutrino and antineutrino mode 
analyses \cite{mb_lowe,mb_osc_anti}. The detector cannot distinguish
between neutrino and antineutrino interactions 
on an event-by-event basis. To help 
constrain the $\bar \nu_e/\nu_e$ candidate events, a
$\bar \nu_\mu/\nu_\mu$ sample is formed. The separation of $\nu_\mu$ 
from $\bar \nu_\mu$ in this large CCQE sample is accomplished by fitting the 
observed angular distribution of the outgoing muons to a linear combination 
of the differing CCQE angular distributions for final state 
$\mu^+$ and $\mu^-$.  
Relative to the Monte Carlo prediction, the $\mu^+$ yield required an increase 
of 1.20 to the rate of $\pi^-$ decays $(\bar \nu_\mu$), while the $\mu^-$ 
yield is 0.99 of its predicted rate. Overall, the normalization required a 
13\% increase, which is compatible with the combined neutrino flux and cross 
section uncertainties \cite{nubarLOI}. A sample of 24,771 data events pass the 
$\bar \nu_\mu$ CCQE selection requirements. The neutrino and antineutrino 
content of the sample are 22\% and 78\%, respectively.

The oscillation parameters are extracted from a combined fit to the 
$\bar \nu_e/\nu_e$
CCQE and $\bar \nu_\mu/\nu_\mu$ CCQE event distributions. Any possible 
$\bar \nu_\mu \rightarrow \bar \nu_e$ signal, as well as some $\bar \nu_e$ 
backgrounds interact through a similar process as $\bar \nu_\mu$ 
CCQE events and are additionally 
related to the $\bar \nu_\mu$ CCQE events through the 
same $\pi^-$ decay chain at production. These correlations enter through the 
off-diagonal elements of the covariance matrix used in the $\chi^2$ 
calculation, relating the contents of the bins of the $\bar \nu_e$ CCQE and 
$\bar \nu_\mu$ CCQE distribution. This procedure maximizes the sensitivity 
to $\bar \nu_\mu \rightarrow \bar \nu_e$ oscillations when systematic 
uncertainties are included \cite{Schmitz:2008zz}.

\begin{table}[t]
\vspace{-0.1in}
\caption{\label{signal_bkgd} \em The expected (unconstrained) number of events
for different $E_\nu^{QE}$ ranges from all of the backgrounds in the $\bar{\nu}_e$ appearance analysis and for the LSND expectation (0.26\% oscillation probability averaged over neutrino energy) 
of $\bar{\nu}_{\mu}\rightarrow\bar{\nu}_e$ oscillations, for $5.66\times10^{20}$ POT.
}
\small
\begin{ruledtabular}
\begin{tabular}{ccc}
Process&$200-475$ MeV&$475-1250$ MeV \\
\hline
$\nu_\mu$ \& $\bar \nu_\mu$ CCQE & 4.3 & 2.0 \\
NC $\pi^0$ & 41.6 & 12.6 \\
NC $\Delta \rightarrow N \gamma$ & 12.4 & 3.4\\
External Events & 6.2 & 2.6 \\
Other $\nu_\mu$ \& $\bar \nu_\mu$ & 7.1 & 4.2 \\
\hline
$\nu_e$ \& $\bar \nu_e$ from $\mu^{\pm}$ Decay & 13.5 & 31.4 \\
$\nu_e$ \& $\bar \nu_e$ from $K^{\pm}$ Decay & 8.2 & 18.6 \\
$\nu_e$ \& $\bar \nu_e$ from $K^0_L$ Decay & 5.1 & 21.2 \\
Other $\nu_e$ \& $\bar \nu_e$ & 1.3 & 2.1 \\
\hline
Total Background &99.5&98.1 \\
0.26\% $\bar{\nu}_{\mu}\rightarrow\bar{\nu}_e$ & 9.1 & 29.1\\
\end{tabular}
\vspace{-0.2in}
\end{ruledtabular}
\normalsize
\end{table}

The number of predicted $\bar{\nu}_e$ CCQE background events for 
different ranges of reconstructed neutrino energy ($E_\nu^{QE}$)
is shown in Table \ref{signal_bkgd}. 
The background estimates include both antineutrino and neutrino events, the latter representing 44\% of the total background. 
The predicted backgrounds to the $\bar{\nu}_e$ CCQE sample are constrained by 
measurements at MiniBooNE 
and include neutral current (NC) $\pi^{0}$ events \cite{mb_pi0}, 
$\Delta\rightarrow N\gamma$ radiative decays, and external
events from neutrino interactions outside the detector. 
Other backgrounds from mis-identified $\nu_{\mu}$ or $\bar{\nu}_{\mu}$ 
\cite{mb_numuccqe,mb_numuccpi} and from
intrinsic $\nu_e$ and $\bar{\nu}_e$ events from the $\pi\rightarrow\mu$ decay chain
receive the $\bar{\nu}_{\mu}$ CCQE normalization correction according to their 
parentage at 
production ($\pi^+$ or $\pi^-$). 

\begin{figure}[tbp]
%\vspace{-0.1in}\centerline{\includegraphics[angle=0, width=9.0cm, trim=0.0 0.0 0.0 0.0]{prl200_stack_excess_wbf.eps}}
\vspace{-0.1in}\centerline{\includegraphics[angle=0, width=9.0cm, trim=0.0 0.0 0.0 0.0]{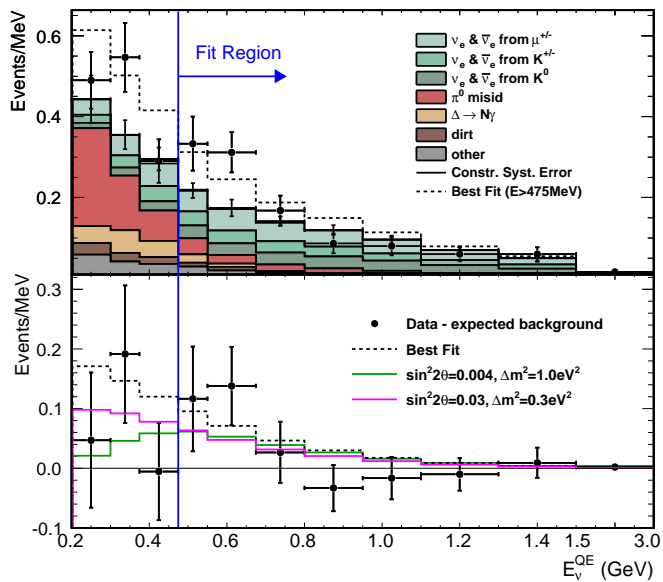}}
\vspace{0.1in}
\caption{Top: The $E_\nu^{QE}$ distribution 
for $\bar{\nu}_e$ CCQE data (points with statistical errors) and background (histogram with systematic errors). Bottom: The event excess as a function of $E_\nu^{QE}$. Also shown are the expectations from the best oscillation fit 
with $E_\nu^{QE}>475$ MeV, $(\Delta m^2, \sin^2 2 \theta)$ = (0.064 eV$^2$, 0.96), where the fit is extrapolated below 475 MeV, 
and from two other oscillation parameter sets in the allowed region. No
correction has been made for the low-energy excess of events
seen in neutrino mode below 475 MeV. All known systematic errors are included 
in the systematic error estimate. }
\label{excess}
\vspace{-0.2in}
\end{figure}

If the low-energy excess observed
during neutrino-mode running \cite{mb_osc} were scaled 
by the total neutrino flux,
the expected excess for antineutrino mode running would be 12 events
for $200<E_\nu^{QE}<475$ MeV. (These events are not included in Tables 
\ref{signal_bkgd} and \ref{signal_bkgd3} and in Fig. \ref{excess} because
they occur below 475 MeV and the origin of these events is unexplained.)

Systematic uncertainties are determined by considering the predicted
effects on the $\bar{\nu}_{\mu}$ and $\bar{\nu}_e$ CCQE rate 
from variations of actual parameters.
These include uncertainties in the neutrino and antineutrino flux estimates, 
uncertainties in neutrino cross sections, most of which are determined by in
situ cross-section 
measurements at MiniBooNE, and uncertainties in 
detector modeling and reconstruction. 
By considering the variation from each source of systematic uncertainty on the $\bar{\nu}_e$ CCQE signal, background, and $\bar{\nu}_{\mu}$ CCQE prediction as a function of $E_{\nu}^{QE}$, a covariance matrix in bins of $E^{QE}_{\nu}$ is constructed. 
This matrix includes correlations between $\bar{\nu}_e$ CCQE (signal and background) and $\bar{\nu}_{\mu}$ CCQE and is used in the $\chi^2$ calculation of the oscillation fit.

\begin{table}[b]
\vspace{-0.2in}
\caption{\label{signal_bkgd3} \em The number of data, fitted (constrained)
background, and excess events in the $\bar{\nu}_e$ analysis for different
$E_\nu^{QE}$ ranges. The uncertainties include both statistical and constrained 
systematic errors. All known systematic errors are included
in the systematic error estimate.}
\begin{ruledtabular}
\begin{tabular}{cccc}
$E_\nu^{QE}$ Range&Data&Background&Excess \\
\hline
$200-475$ MeV&119&$100.5 \pm 10.0 \pm 10.2$&$18.5 \pm 14.3$ \\
$475-675$ MeV&64&$38.3 \pm 6.2 \pm 3.7$&$25.7 \pm 7.2$ \\
$475-1250$ MeV&120&$99.1 \pm 10.0 \pm 9.8$&$20.9 \pm 14.0$ \\
$475-3000$ MeV&158&$133.3 \pm 11.5 \pm 13.8$&$24.7 \pm 18.0$ \\
$200-3000$ MeV&277&$233.8 \pm 15.3 \pm 16.5$&$43.2 \pm 22.5$ \\
\end{tabular}
\end{ruledtabular}
\end{table}

Fig. \ref{excess} (top) shows the $E_\nu^{QE}$ distribution for 
$\bar{\nu}_e$ CCQE observed data and background. A total of 277 events pass 
the $\bar{\nu}_e$ event selection requirements with $200<E_\nu^{QE}<3000$ MeV, 
compared to an expectation of $233.8 \pm 15.3 \pm 16.5$ events, where the 
uncertainty includes both statistical and systematic errors, respectively. 
This corresponds to an excess of $43.2 \pm 22.5$ events. 
(In the previous neutrino run analysis, event totals 
were considered in two energy regions: 200 - 475 MeV 
and 475 - 3000 MeV, where the latter region was the energy range for the 
neutrino oscillation search.  For the antineutrino data,
the excess for $475<E_\nu^{QE}<3000$ MeV is $24.7 \pm 18.0$ events.)
In the 
%oscillation-sensitive 
energy range from $475 < E_\nu^{QE} < 1250$ MeV, the 
observed $\bar \nu_e$ events, when constrained by the $\bar \nu_\mu$ data 
events, have a $\chi^2/DF = 18.5/6$ and a probability of 
0.5\% for a background-only hypothesis. (This compares to the 40\% 
probability that is observed in neutrino mode \cite{mb_lowe} for the 
same energy range.)
DF is the effective number of degrees of 
freedom from frequentist studies. 
%For the 
%combined $\bar \nu_\mu$ and $\bar \nu_e$ data in the energy range $475 < 
%E_\nu^{QE} < 3000$ MeV, the background-only hypothesis 
%yields a $\chi^2/DF = 26.8/14.9$ with a $\chi^2$-probability 
%of 3.0\%. 
%This is higher than 
%the 0.5\% probability found for the $\bar \nu_e$ signal region due to the 
%inclusion of $\bar \nu_\mu$ and higher-energy $\bar \nu_e$ bins, where no 
%signal is expected.
The number of data, fitted background, and excess events for different 
$E_\nu^{QE}$ ranges are summarized in Table \ref{signal_bkgd3}.

\begin{figure}[tbp]
\vspace{-0.2in}
\centerline{\includegraphics[angle=-90,width=7cm, trim=112 128 96 104]{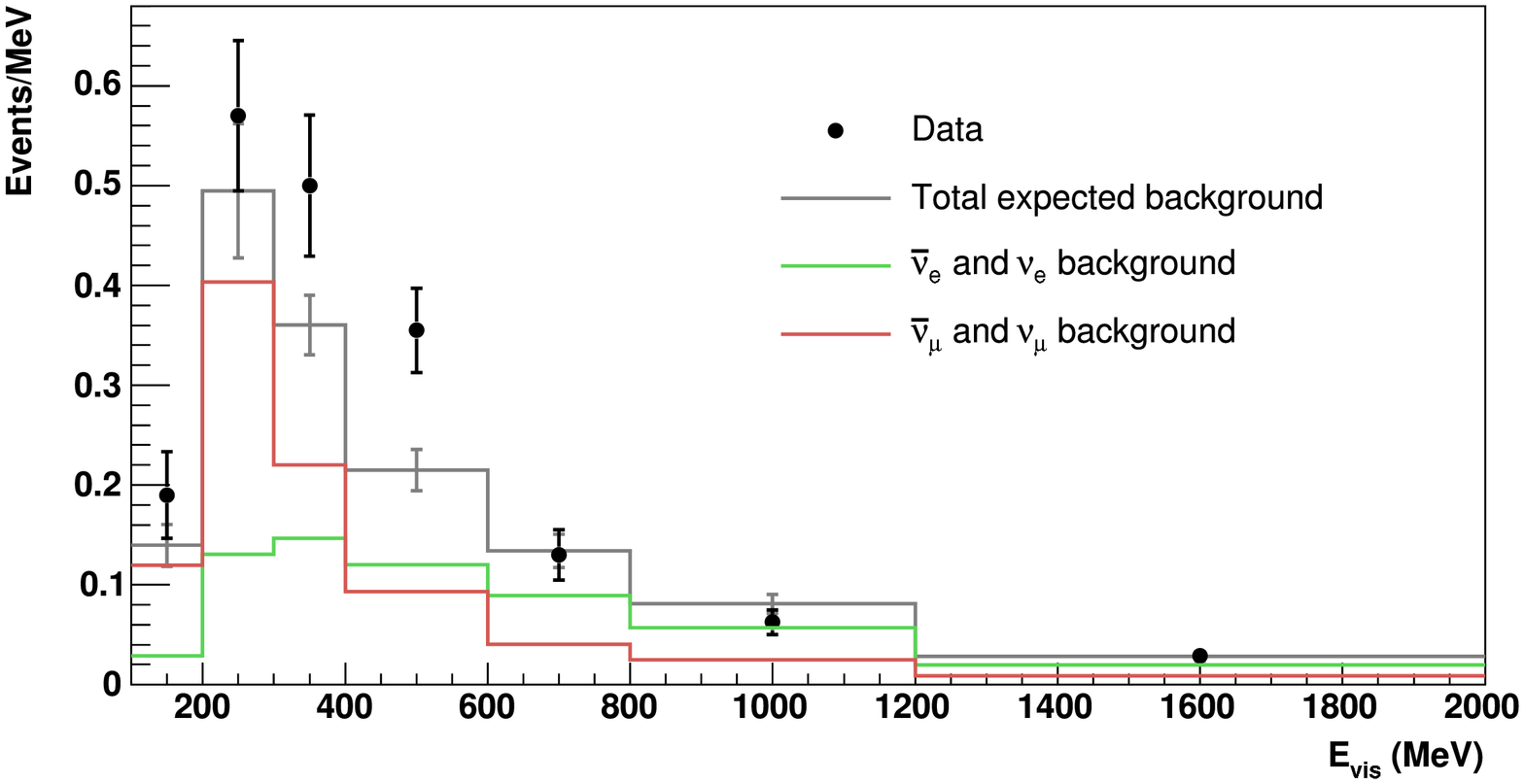}}
\centerline{\includegraphics[angle=-90,width=7cm, trim=88 128 64 104]{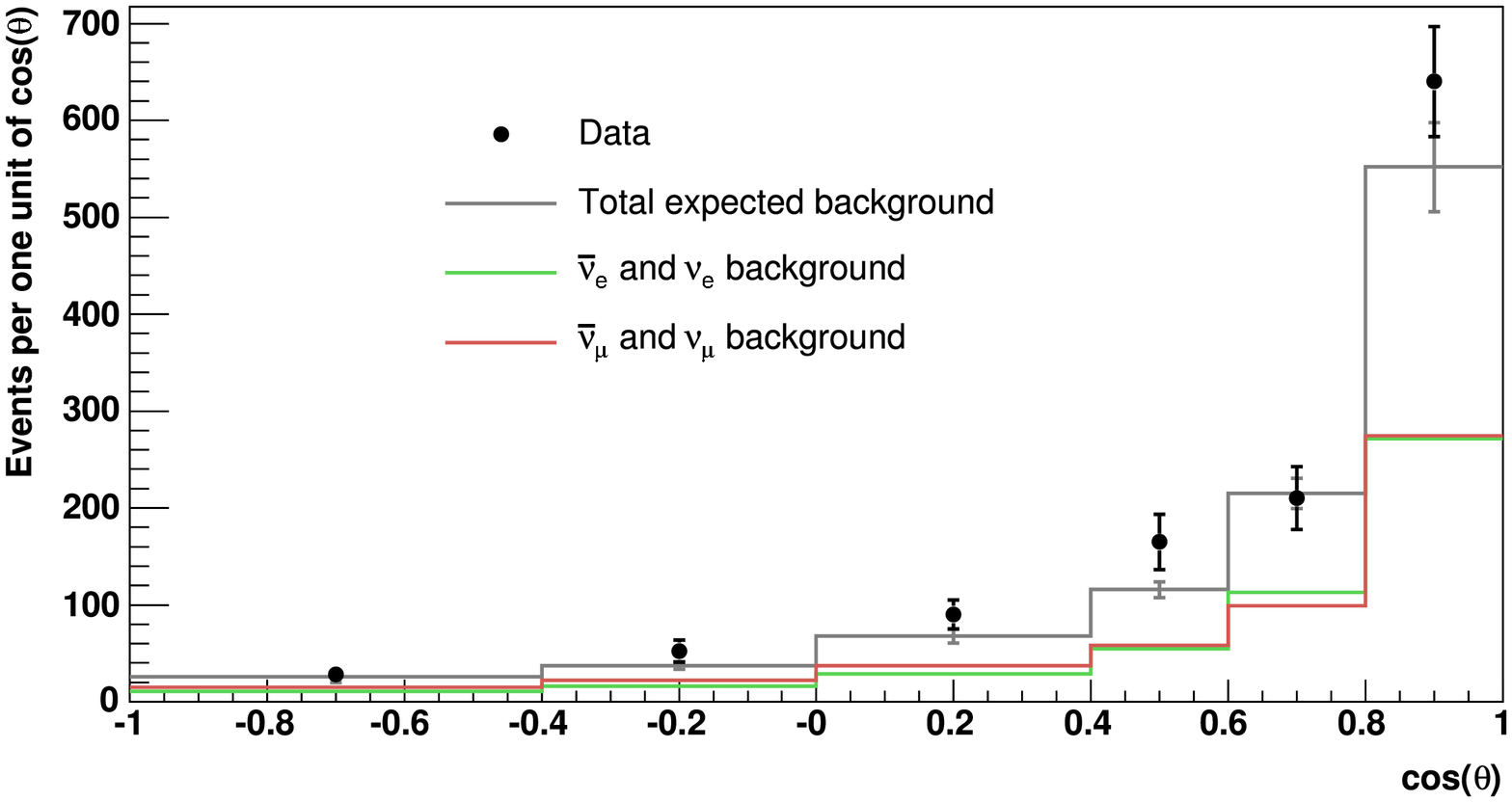}}
\vspace{0.2in}
\caption{The $E_{vis}$ (top panel) and $\cos(\theta)$ (bottom panel) distributions for data (points with statistical errors) and backgrounds (histogram with systematic errors) for $E_\nu^{QE} > 200$ MeV.} 
\label{data_mc4}
\vspace{-0.2in}
\end{figure}

Fig.~\ref{data_mc4} shows the observed and predicted event distributions as functions of reconstructed $E_{vis}$  and $\cos (\theta)$ for $200 < E_\nu^{QE} < 3000$ MeV. 
$E_{vis}$ is the measured visible energy, while $\theta$ is the scattering angle of
the reconstructed electron with respect to the incident neutrino direction.
The background-only $\chi^2$ values for the $\bar \nu_e$ and $\bar \nu_\mu$
data are 
$\chi^2/DF=23.8/13$ and $\chi^2/DF=13.6/11$ for $E_{vis}$ and $\cos (\theta)$, 
respectively. 
%Therefore, the $E_{vis}$ distribution does not match well the
%shape of the estimated backgrounds, while the $\cos (\theta)$
%distribution matches the background shape.

Many checks have been performed on the data to ensure that the backgrounds 
are estimated correctly.
Beam and detector stability checks show that the neutrino event rate is stable to $<2\%$ and
that the detector energy response is stable to $<1\%$. In addition, the fractions of neutrino
and antineutrino events are stable over energy 
and time, and the inferred external event rates are similar
in both neutrino and antineutrino modes. Furthermore,
any single background would have to be increased by more than
3 $\sigma$ to explain the observed excess of events.
An additional check comes from the data in neutrino mode, which has a similar
background to antineutrino mode and where good agreement 
is obtained between the data and Monte Carlo simulation for $E_\nu^{QE} > 475$ MeV. As a final check, 
the event rate of candidate $\bar \nu_e$ events in the last 
$2.27 \times 10^{20}$ POT is found to be $1.9 \sigma$ higher than the 
candidate event rate in the first $3.39 \times 10^{20}$ POT \cite{mb_osc_anti};
however, the $\bar \nu_\mu$ event rates are found to be similar for the two 
running periods.

\begin{figure}[tbp]
\centerline{\includegraphics[width=9.0cm]{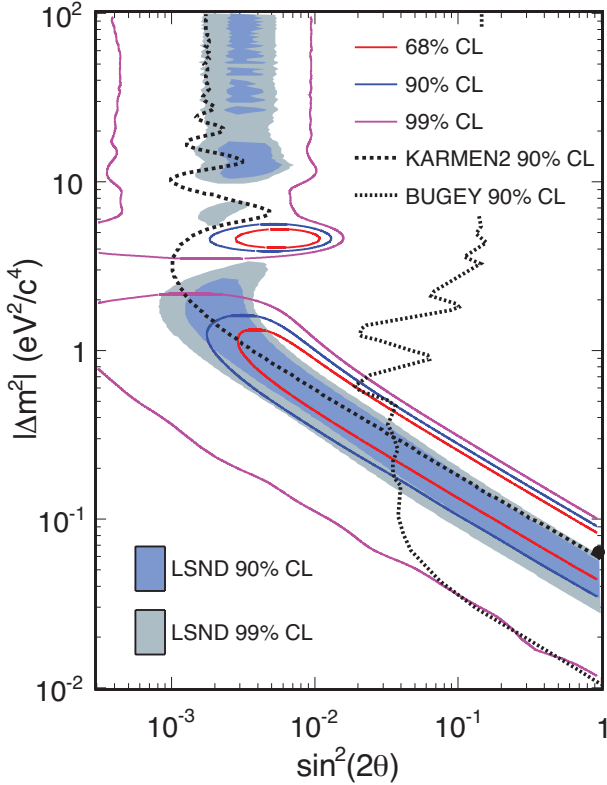}}
\caption{MiniBooNE 68\%, 90\%, and 99\% C.L. allowed regions for events with 
$E^{QE}_{\nu} > 475$ MeV within a two neutrino $\bar{\nu}_{\mu}\rightarrow\bar{\nu}_e$ oscillation model. Also shown are limits from KARMEN \cite{karmen} and Bugey \cite{bugey}. 
The Bugey curve is a 1-sided limit for $\sin^2 2\theta$ corresponding to $\Delta\chi^2 = 1.64$, 
while the KARMEN curve is a ``unified approach'' 2D contour. 
The shaded areas show the 90\% and 99\% C.L. LSND allowed regions. 
The black dot shows the best fit point.}
\label{limit}
\vspace{-0.2in}
\end{figure}

\begin{figure}[tbp]
\centerline{\includegraphics[width=9.0cm]{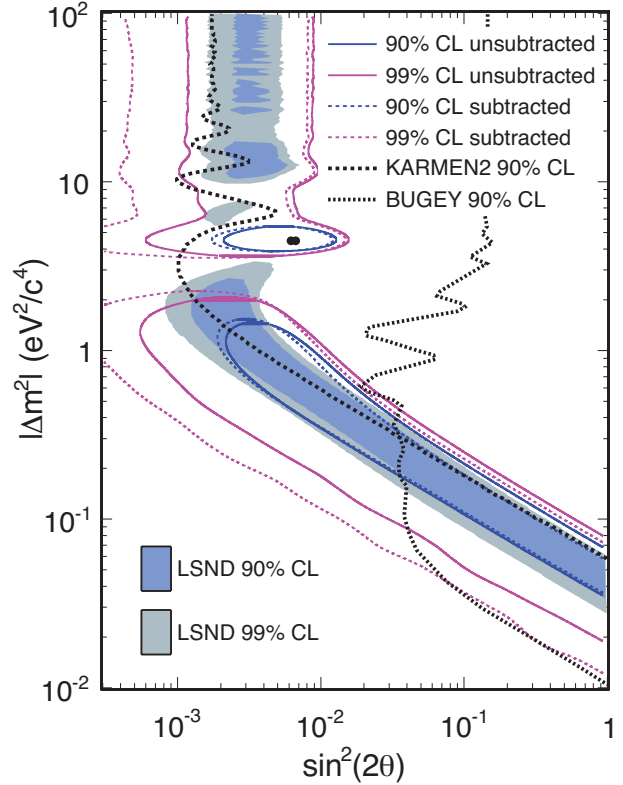}}
\caption{MiniBooNE 90\% and 99\% C.L. allowed regions for events with
$E^{QE}_{\nu} > 200$ MeV within a two neutrino 
$\bar{\nu}_{\mu}\rightarrow\bar{\nu}_e$ oscillation model. 
The solid (dashed) curves are without (with) the subtraction of the
expected 12 event excess in the $200<E_\nu^{QE}<475$ MeV low-energy region
from the neutrino component of the beam.
Also shown are limits from KARMEN \cite{karmen} and Bugey \cite{bugey}.
The shaded areas show the 90\% and 99\% C.L. LSND allowed 
regions. The black dots show the best fit points.}\label{limit2}
\vspace{-0.2in}
\end{figure}

Fig.~\ref{excess} (bottom) shows the event excess as a function of $E_\nu^{QE}$.
Using a likelihood-ratio technique,
the best MiniBooNE oscillation fit for $475<E_\nu^{QE}<3000$ MeV occurs at
($\Delta m^2$, $\sin^22\theta$) $=$ (0.064 eV$^2$, 0.96). 
%and has a $\chi^2$ of 16.4 for 12.6 DF,
%corresponding to a $\chi^2$-probability of 20.5\%. 
The energy range $E^{QE}_{\nu} > 475$ MeV has been chosen
for the fit as this is the energy range MiniBooNE used for
searching for neutrino oscillations. Also, this energy range avoids the
region of the unexplained low-energy excess in neutrino mode \cite{mb_lowe}.
This best-fit point has only a slightly lower $\chi^2$ than other points 
in the allowed band. 
%(The points at ($\Delta m^2$, $\sin^22\theta$) $=$ (1.0 eV$^2$, 0.004) 
%and (0.3 eV$^2$, 0.03) have probabilities that are 39\% and 54\%,
%respectively, of the best-fit point.) 
The $\chi^2$ for the best-fit point in the
%oscillation-sensitive 
energy range of $475<E_\nu^{QE}<1250$ MeV
is 8.0 for 4 DF, corresponding to a $\chi^2$-probability of 8.7\%.
The probability of the background-only fit relative to the best
oscillation fit is 0.6\%.
Fig.~\ref{limit} shows the MiniBooNE
68\%, 90\%, and 99\% C.L. closed contours for 
$\bar{\nu}_{\mu}\rightarrow\bar{\nu}_e$ oscillations in the 
$475<E_\nu^{QE}<3000$ MeV energy range, where 
frequentist studies were performed
to determine the C.L. regions.
The allowed regions are in agreement with the LSND allowed region.
%These allowed regions appear to be different from the allowed
%regions for $\nu_\mu \rightarrow \nu_e$ oscillations \cite{mb_osc}.
The MiniBooNE closed contours for
$\bar{\nu}_{\mu}\rightarrow\bar{\nu}_e$ oscillations in the 
$200<E_\nu^{QE}<3000$ MeV energy range are similar, as shown in 
Fig.~\ref{limit2}. The solid (dashed) curves are without (with) the 
subtraction of the expected 12 event excess in the $200<E_\nu^{QE}<475$ MeV 
low-energy region from the neutrino component of the beam. 
The best oscillation fits without and with this subtraction occur at
($\Delta m^2$, $\sin^22\theta$) $=$ (4.42 eV$^2$, 0.0066) and 
(4.42 eV$^2$, 0.0061), respectively, while the corresponding
$\chi^2$-probabilities in the $200<E_\nu^{QE}<1250$ MeV energy range
are 10.9\% and 7.5\%.

\begin{figure}[tbp]
%\vspace{-0.1in}\centerline{\includegraphics[angle=0, width=11.0cm, trim=0.0 0.0 0.0 0.0]{LSNDMiniBooNEOscData-NueBarExpanded.eps}}
%\vspace{-0.1in}\centerline{\includegraphics[angle=0, width=10.5cm, trim=0.0 100.0 0.0 0.0]{prl-plot-v2.eps}}
\vspace{-0.1in}\centerline{\includegraphics[angle=0, width=10.5cm, trim=0.0 100.0 0.0 0.0]{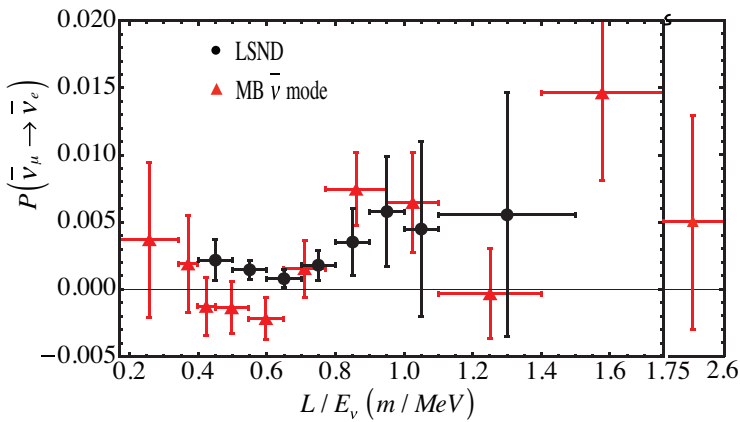}}
\vspace{0.1in}
\caption{The oscillation
probability as a function of $L/E_\nu^{QE}$ for $\bar \nu_\mu \rightarrow
\bar \nu_e$ candidate events from MiniBooNE and LSND. The data points include
both statistical and systematic errors.}
\label{L_E}
\vspace{-0.2in}
\end{figure}

A further comparison between the MiniBooNE and LSND antineutrino data sets
is given in Fig. \ref{L_E}, which shows the oscillation
probability as a function of $L/E_\nu$ for $\bar \nu_\mu \rightarrow 
\bar \nu_e$ candidate events in the $L/E_\nu$ range where
MiniBooNE and LSND overlap. The data used for LSND and MiniBooNE correspond
to $20<E_\nu<60$ MeV and $200<E_\nu^{QE}<3000$ MeV, respectively. 
The oscillation probability is defined as the
event excess divided by the number of events expected for 100\% 
$\bar \nu_\mu \rightarrow \bar \nu_e$ transmutation, while $L$
is the reconstructed distance travelled by the antineutrino from the
mean neutrino production point to the interaction vertex and
$E_\nu$ is the reconstructed antineutrino energy. The $L/E_\nu$ 
distributions for the two data sets are consistent.
%The agreement in the $L/E_\nu^{QE}$ distributions for the two data sets
%suggests that they may arise from a similar process.

In summary, the MiniBooNE experiment 
observes an excess of $\bar{\nu}_e$ events in the energy region 
above $E_\nu^{QE}$ of 475 MeV for a data sample corresponding to $5.66\times10^{20}$ POT. 
A model-independent hypothesis test gives a probability 
of 0.5\% for the data to be consistent with the expected 
backgrounds in the 
%oscillation-sensitive 
energy range of $475 < E_\nu^{QE} < 1250$ MeV, 
and a likelihood-ratio fit gives a 0.6\% 
probability for background-only relative to the best 
oscillation fit.
The allowed regions from the fit, shown in Fig. \ref{limit}, 
are consistent with $\bar{\nu}_{\mu}\rightarrow\bar{\nu}_e$ oscillations 
in the 0.1 to 1 eV$^2$ $\Delta m^2$ range 
and consistent with the allowed region reported by the LSND 
experiment \cite{lsnd}. Additional running in antineutrino mode 
is expected to approximately double the current number of POT.
%The allowed regions, however, appear to be different from the allowed
%regions for $\nu_{\mu}\rightarrow \nu_e$ oscillations \cite{mb_osc}.

\begin{acknowledgments}
We acknowledge the support of Fermilab, the Department of Energy,
and the National Science Foundation, and
we acknowledge Los Alamos National Laboratory for LDRD funding. We also acknowledge the use of Condor software for the analysis of the data.
\end{acknowledgments}

% Appendix
%\appendix
%\section{Appendices}

%\newpage %Just because of unusual number of tables stacked at end
\bibliography{prl}% Produces the bibliography via BibTeX.

\end{document}